\documentclass[12pt,fullpage]{article}
\usepackage{fullpage}
\usepackage{url}
\begin{document}
\title{\bf SoK: Scope and Mission of CS\&Law}
\author{
Joan Feigenbaum\\
Yale University\\
New Haven, CT 06520, USA\\
{\tt joan.feigenbaum@yale.edu}
\and
Daniel J. Weitzner\\
Massachusetts Institute of Technology\\
Cambridge, MA 02139, USA\\
{\tt weitzner@mit.edu}
}
\date{October 8, 2025}

\maketitle
\begin{abstract}
  We systematize the intellectual scope of the ACM Computer Science and Law Symposium (CS\&Law). In particular, we address the meaning and importance of the word “and” in the name of the symposium.  We identify previously published papers (from CS\&Law and other forums) that exemplify different aspects of the CS\&Law scope and note that the scope is expected to evolve as the symposium and the community grow and change.  To round out our systematization of the still nascent research area, we also discuss the mission of CS\&Law: What might the symposium seek to accomplish beyond providing a forum for intellectual exchange and community formation? 
\end{abstract}

\section{Introduction}\label{sec-introduction}
Advances in artificial intelligence, cryptography, algorithmic data processing, networking, databases, software engineering, and many other fields of computer science have raised difficult legal questions. At the same time, computational methods have opened new perspectives on legal problems in domains ranging from criminal procedure to evidence to intellectual property. These intersections have created a growing need for research that combines deep understanding of the power and the limitations of computing with expertise in multiple aspects of law.

The ACM Symposium on Computer Science and Law (CS\&Law) is the flagship conference for the emerging field of computer science and law. It brings together a community of scholars, lawyers, and computing professionals who are fluent in both computational thinking (with its rigorous mathematical formalisms) and legal scholarship and thought (with its equally rigorous yet human-centric set of principles, methodologies, and goals). Central to the study of “computer science and law” is the creation of a body of scholarship aimed at the co-design of law and computing technology that promotes sociotechnical goals.

This paper explores the scope and mission of CS\&Law and establishes boundaries between it and some adjacent conferences.  While continuing to adhere to some basic principles, the scope will probably grow over time.  By contrast, the mission will be more enduring.

Our intended audiences include:
\begin{itemize}
    \item {\bf The current CS\&Law community}: What is the community trying to accomplish, and how might it grow and improve?
    \item {\bf Prospective community members}: Is this a forum in which they might wish to participate?
    \item {\bf Academic department chairs and tenure-and-promotion committees}: How might tenure and promotion dossiers successfully present the content and scholarly practices of “computer science and law” and make the case that a candidate has earned tenure or promotion?
    \item {\bf Funding sources}: What are the compelling future directions of the CS\&Law community and why should its work be supported?
\end{itemize}

\subsection{The Intersection Rule}\label{subsec-intersection}

The defining characteristic of CS\&Law is true interdisciplinarity.  This subsection examines both what interdisciplinarity means in this context and why it is essential.

\subsubsection{What is the intersection rule?}\label{subsubsec-what}
The most important thing to keep in mind when understanding whether a paper meets the requirements of CS\&Law is that “and” means “logical and”\ --\  {\it intersection, not union}. The relevant text in the Call for Papers\footnote{This quote is taken from the 2026 version of the CfP.} states that the CS\&Law organizers solicit “submissions that take an interdisciplinary approach to the field of computer science and law. [They] seek papers in the intersection of the two fields that combine rigorous, technical computer-science reasoning with rigorous legal analysis. Submissions should address a problem of interest to both fields, engage with the relevant literature from both, and integrate the two disciplines in a way that surpasses what either could accomplish by itself.”  

Insistence on a literal interpretation of the word “and” is not a demand that CS\&Law researchers do something that has never been done before or that they clear an impossibly high bar.  Before the inaugural CS\&Law symposium was held in 2019, there were a number of papers published that satisfy the intersection rule. A small sample of the entire set includes papers on legal protection of software~\cite{Samuelson1994}, law-enforcement access to encrypted material~\cite{Abelson1997,Abelson2015}, and the tension between rules and principles in technology-policy debates~\cite{Feigenbaum2018}.  CS\&Law was established in order to create an official home for this type of research and encourage more people to do it.

Similarly, the strict definition of the CS\&Law scope is not an implicit criticism of related papers that fall outside of the scope: Computer-science papers that apply computing techniques to law and law papers that apply legal analysis to computer-related technologies are obviously of interest to people in the CS\&Law community.  However, a number of high-quality, disciplinary forums in which to publish such papers had already been established by the time CS\&Law’19 was convened, and they are still the appropriate venues for such papers. Those forums include (but are not limited to) the law-and-technology journals produced by influential law schools, the Privacy Law Scholars Conference (PLSC), the International Conference on Legal Knowledge and Information Systems (JURIX), and annual computer-science conferences in various research areas that implicate law and policy, such as machine learning ({\it e.g.}, the Conference on Neural Information Processing Systems [NeurIPS] and the International Conference on Machine Learning [ICML]), computer and information security ({\it e.g.}, the IEEE Symposium on Security and Privacy [Oakland], the ACM Conference on Computer and Communications Security [CCS], and the USENIX Security Symposium), and human-computer interaction ({\it e.g.}, the ACM Conference on Human Factors in Computing Systems [CHI], the ACM Symposium on User Interface Software and Technology [UIST], and the ACM Conference on Computer-Supported Cooperative Work \& Social Computing [CSCW]).  In contrast to these disciplinary venues, CS\&Law seeks to increase the depth, breadth, visibility, and status of research in the intersection of computer science and law that is {\it necessarily interdisciplinary} in that it relies upon the investigators’ ability to combine knowledge of the two fields in a substantial way.

\subsubsection{Why is the intersection rule essential?}\label{subsubsec-why}
The previous section explains why a paper that is suitable for a disciplinary venue (in computer science or law) may not be suitable for CS\&Law.  What about the other way around?  In order to be accepted by CS\&Law, must a submission be a good candidate for acceptance to an adjacent disciplinary venue?

The answer is no, and the reasoning that leads to that answer reveals the value of the intersection rule.  Although many CS\&Law papers present results that {\it would} be accepted for publication in high-quality disciplinary venues, a paper is not {\it required} to do so in order to be accepted to CS\&Law.  The problem addressed in a CS\&Law paper may be one in which both the computational challenges and the legal challenges have known solutions but in which it is unclear whether the solutions are compatible.  Determining whether they are compatible and, if so, how they can be combined may be the paper’s novel content ({\it i.e.}, the “research”) from the point of view of CS\&Law.  Similarly, a paper that contains a modest amount of novel computer-science content or novel legal content (but not enough to warrant publication in a disciplinary conference or journal) may be suitable for CS\&Law if significant intellectual effort is needed to combine the two aspects of the solution. 

The potential for the ``and'' to {\it be} the research content of a paper, as opposed to merely the locus of the research content, can be understood by considering a particular open problem\ --\ {\it e.g.}, law-enforcement access to encrypted material.\footnote{In this discussion, the word ``access'' refers to legal, ``front-door'' access of the type advocated by many people in law enforcement, including James Comey when he was FBI Director\ --\ not to ``magic back-door access'' of the type that gray-hat hackers can often provide}. As noted in the previous section, it is a problem that clearly fits into the CS\&Law scope.  There has been some excellent work on it, but it has not been definitively solved.  Would a definitive
solution require substantial research in either law or computer science of sufficient novelty and difficulty to warrant publication in a high-quality disciplinary venue?

Not necessarily.  From the standpoint of legal requirements alone, one could say that the problem is solved: If a law-enforcement officer has a properly authorized warrant to search a digital device or system believed to hold relevant material, then he is {\it legally} entitled to conduct the search.  However, in both digital and analog realms, legal authorization to search does not guarantee that the search will succeed; while end-to-end encryption mechanisms in which decryption is entirely under the control of end users are legal, as they currently are in the US and many other countries, the officer may be {\it technically} unable to access the material he seeks in plaintext form.  Whether law-enforcement access protocols could be added to the computing and communication infrastructure in such a way that they enable {\it only} properly authorized access, do not prevent the widespread use of strong encryption (which is currently users' best defense against identity theft, intellectual-property theft, and other online crimes), and do not impose unacceptable costs is an open technical question.  That the design and implementation of such protocols would be at best extremely difficult has been  argued by numerous experts, including the authors of \cite{Abelson1997,Abelson2015}, but no one has proven that it is impossible. Thus, the feasibility of law-enforcement access is an open question from the CS\&Law point of view. In response to this earlier CS\&Law scholarship, Australia has imposed a so-called technical-assistance requirement to compel service providers to enable law enforcement to decrypt data held by those service providers but also added a caveat that such assistance is not required if it would lead to systemic vulnerabilities or otherwise be unreasonable.\footnote{The Assistance and Access Act 2018, Australia} Hence, a new round of CS\&Law work is now required to determine how to evaluate such terms in the law, from both legal \textit{and} computer-science perspectives.

For researchers to provide a definitive answer to the question of whether law-enforcement access to encrypted material is technically feasible, they would need to define and scope it precisely 
and either provide a solution that is provably secure and efficient or prove that no such solution exists.  That project would lie squarely in the ``and'' of CS\&Law, requiring the interplay of rigorous legal analysis and rigorous computer science.  It {\it might} require substantial new legal scholarship or new work in cryptography, information theory, system architecture, and/or other computer-science areas, but it might not.  If it did, then it would make a good disciplinary paper, but either way it would make a good CS\&Law paper.

\subsection{Adjacent Interdisciplinary Conferences}\label{subsec-adjacent}

Computer-science researchers have engaged in sociotechnical research for at least 30 years.\footnote{Computer-science researchers played a major role in the “crypto wars” of the 1990s.}  Recently, the interplay of computer science and various social-science disciplines has grown considerably in volume, visibility, and acceptance by institutional leaders.  A number of interdisciplinary conferences have been established that are similar to CS\&Law in some ways and attract some of the same people.  Among the most successful are the ACM Conference on Economics and Computation (EC), related Econ-oriented venues such as the International Symposium on Algorithmic Game Theory (SAGT) and the Conference on Web and INternet Economics (WINE), the ACM Conference on Fairness, Accountability, and Transparency (FAccT), and the Symposium on Foundations of Responsible Computing (FORC).

Legal, policy, and regulatory issues play a role in some of the papers at each of these conferences.  CS\&Law is unique, however, in being devoted exclusively to research in the intersection of computer science and law. 

\subsection{Academic Careers}\label{subsec-academic}

Since the beginning of its existence as a distinct research community, CS\&Law has been fortunate to count a number of prominent senior academics and junior rising stars as members.  We hope and expect that the academic part of our community will continue to grow both in size and in influence.  Tenure cases are often harder to make in new fields and in interdisciplinary fields, and we encourage department chairs and tenure-committee chairs to keep the following points in mind when making cases for people in CS\&Law.
\begin{itemize}
    \item {\bf Field description}: General descriptions of the intellectual content of the research area and its constituent topics can be found here and in Calls for Papers for the CS\&Law symposium.
    \item {\bf Scholarly visibility}: Many tenure-track CS\&Law scholars will have published in at least one premier disciplinary forum in CS or Law (including but not limited to those listed in §\ref{subsec-intersection}) as well as in the CS\&Law symposium. Prominent, senior computer scientists and legal scholars in the CS\&Law world can write letters and have experience doing so, but letter writers from relevant disciplinary areas can also be used.
    \item {\bf Student interest}: Demand for interdisciplinary courses that involve CS, especially AI, is growing, and CS\&Law is especially popular among undergrads at elite colleges. 
\end{itemize}

\section{Illustrative Papers}\label{section-illustrative}

This section identifies papers that serve as exemplars of the intersection rule. We have drawn them both from the proceedings of previous CS\&Law symposia and from other forums. They are illustrative examples and {\it are} \textit{\textbf{not}} {\it intended as an exhaustive list}.

Two ways in which a research project that satisfies the intersection rule can be conceived suggest themselves very naturally.  The researchers could start from a legal perspective: With respect to a particular law or regulation, it may be unclear whether relevant technologies comply with it or how one could demonstrate that they do. Conversely, the researchers could start from a technological perspective: With respect to a particular technology or its use in a particular application domain, it may be unclear whether it complies with applicable laws and regulations or how to demonstrate that it does. We start by providing examples of papers that address conceptual and practical gaps of this sort. 

\newpage

\begin{enumerate}
\item[$\bullet$] Starting from a legal perspective
\begin{enumerate}
\item[$\circ$] 
\underline{Paper}: ``Towards formalizing the GDPR’s notion of singling out,'' by Aloni Cohen and Kobbi Nissim~\cite{Cohen2020}

\underline{Relevant law(s) or regulation(s)}: EU General Data Protection Regulation (GDPR)

\underline{Summary}: The GDPR requires any data-release mechanism that purports to render personal data anonymous to ``prevent singling out.''  Although the notion of singling out makes intuitive sense to many people, it is not defined precisely in the text of the GDPR.  Technology providers are thus burdened with uncertainty about whether their offerings conform with the regulation.  The authors close this gap between legal and mathematical thinking by rigorously defining the act of {\it predicate singling out}, offering {\it security against predicate singling out (PSO security)} as the technical standard that data-release mechanisms should meet in order to satisfy GDPR requirements, and proving theorems about PSO security that have legal consequences.

\underline{Value of \&}: Translating legal language into mathematical and computational frameworks is a standard way to practice the intersection rule.  In this paper, it leads to interesting theorems, and that alone is sufficient motivation for the work. However, the paper goes on to provide practical guidance by showing that any differentially private data-release mechanism is PSO secure but that a $k$-anonymous mechanism need not be.  The latter result is especially useful: Before Cohen and Nissim's paper appeared, guidance by the Article 29 Data Protection Working Party, an EU advisory body, claimed that $k$-anonymity eliminated the risk of singling out.
\\

\item[$\circ$] 
\underline{Paper}: ``Age Verification Systems
Will Be a Personal Identifiable Information Nightmare,'' by Sarah Scheffler~\cite{Scheffler2024}

\underline{Relevant law(s) or regulation(s)}: A list of pending and active state age-verification laws is available at \url{https://bit.ly/4e4cZL8}.

\underline{Summary}: In an effort to bar children from websites that may harm them, several US states have instituted or proposed strict online age-verification requirements for various sites, including but not limited to porn sites.  The author explains why online age verification is extremely difficult technically, why requiring websites to do it will severely threaten privacy and security, how such requirements incentivize growth in the nascent ID-verification-service industry (itself a threat to privacy and security), and which alternative approaches to protecting children online may be preferable.

\underline{Value of \&}: Laws requiring online age verification perfectly exemplify a recurring type of clash between politicians and technologists.  The former, often in cooperation with advocacy groups, attempt to solve a genuine social problem by commanding that technologists ``just do it,'' refusing to acknowledge that ``it'' may be technically infeasible, unacceptably costly, or likely to cause worse social problems than the one they are trying to combat.  By explicating (clearly, in layman's terms) the harmful interplay of law and computing technology that age-verification requirements would engender, this paper provides precisely the type of response that is needed when such a clash occurs.
\\

\item[$\circ$] 
\underline{Paper}: ``What constitutes a Deep Fake?
The blurry line between legitimate processing and manipulation under the EU AI Act,'' by Kristof Meding and Christoph Sorge~\cite{Meding2025}

\underline{Relevant law(s) or regulation(s)}: EU AI Act (AIA)

\underline{Summary}: The AIA defines ``deep fakes,'' imposes transparency obligations on providers of technology that generates synthetic content or manipulates existing content, and carves out an exception to one of the obligations for ``AI systems [that] perform an assistive function for standard editing or do not substantially alter the input data provided by the deployer.''  The authors perform a fine-grained evaluation of the language of the AIA that is guided by basic principles and current practice of image processing.  They conclude that deep fakes are ill-defined in the AIA and that it is not clear how a widely deployed, popular editing function like Google’s “best take” can be considered an exception to the transparency obligation. 

\underline{Value of \&}: By combining the principles and practice of image processing with rigorous legal analysis, the authors demonstrate that complying with AIA transparency obligations is difficult for technology providers and deployers.  The stated goal of the work is to foster discussion about what constitutes a deep fake and to raise public awareness of the shortcomings of the AIA as written; we believe that this goal is likely to be met.
\\
\end{enumerate}

In the next five examples, we have identified specific laws for each paper, because they pose particular challenges for the technologies or application domains in question, and the authors deal with them in the papers.  Of course, technologies must comply with {\it all} applicable laws of the countries in which they are deployed.
\item[$\bullet$] Starting from a technological perspective
\begin{enumerate}
\item[$\circ$] 
\underline{Paper}: ``Talkin’ ‘Bout AI Generation: Copyright and the Generative-AI Supply Chain (The Short Version),'' by Katherine Lee, A.~Feder Cooper, and James Grimmlemann~\cite{Lee2024}

\underline{Relevant technology or application}: Generative AI

\underline{Relevant law(s) or regulation(s)}: US copyright law, including but not limited to the fair-use doctrine

\underline{Summary}: The authors grapple with the question of whether and how generative AI may infringe copyright in all of its complexity.  They start from the crucial observation that ``generative AI'' is not a single product but rather an ecosystem of related, interacting technologies.  They go on to explain how and where the ecosystem touches upon many aspects of copyright law, including authorship, similarity, direct and indirect liability, and fair use.  The result of their meticulous analysis of this interplay of technology and law is a holistic intellectual framework called {\it The Generative-AI Supply Chain}, in which upstream technical designs affect downstream uses in essential ways that implicate responsibility for infringement when it occurs. 

\underline{Value of \&}: By breaking generative AI down into eight stages of a process, from creation of works (stage 1) through generation and alignment (stages 7 and 8), the authors expose precisely where companies and users may choose to go down paths that have copyright consequences.  By pinpointing the key decisions that courts will need to make in infringement cases, they identify consequences that can be expected to flow from different liability regimes.  The paper presents an excellent example of research that, in the words of the CS\&Law Call for Papers, ``combine[s] rigorous, technical computer-science reasoning with rigorous legal analysis.''
\\

\item[$\circ$] 
\underline{Paper}: ``Toward Architecture-Driven Interdisciplinary Research: Learnings from a Case Study of COVID-19 Contact Tracing Apps,'' by Fabian Burmeister, Mickey Zar, Tilo B{\"o}hmann, Niva Elkin-Koren, Christian Kurtz, and Wolfgang Schulz~\cite{Burmeister2022}

\underline{Relevant technology or application}: COVID-19 contact-tracing apps

\underline{Relevant law(s) or regulation(s)}: Privacy and data-protection laws in various countries that deployed contract-tracing apps

\underline{Summary}: The authors conduct a novel form of architecture-driven interdisciplinary research at the interface of information systems and law.  They use the data-ecosystem architecture meta-model put forth in~\cite{Burmeister2021} in order to explore the privacy issues and policy
implications of COVID-19 contact-tracing apps.  In particular, they conduct a detailed case study of the HaMagen (``The Shield'') app that launched in Israel in March 2020.  They draw descriptive conclusions about HaMagen, about the architectural meta-model, and about the interdisciplinary research process.

\underline{Value of \&}: The author's detailed explanation of their research methodology demonstrates that the architectural perspective on complex sociotechnical ecosystems is valuable for interdisciplinary research at the interface of information systems and law. In particular, it illustrates how to integrate a legal perspective into the derivation of models and meta-models.
\\

\item[$\circ$]
\underline{Paper}: ``Big Data's Disparate Impact,'' by Solon Barocas and Andrew D. Selbst~\cite{Barocas2016}

\underline{Relevant technology or application}: Use of machine learning and data mining in high-consequence decisions (such as hiring and college admissions) that can involve subjective judgment

\underline{Relevant law(s) or regulation(s)}: US anti-discrimination law, especially the prohibition on discrimination in employment in Title VII of the Civil Rights Act of 1964

\underline{Summary}: The authors start from the well established fact that algorithmic techniques such as machine learning and data mining do not automatically eliminate human biases from decision making, because the data sets that these algorithms work with often inherit the prejudices of prior (human) decision makers. They go on to analyze these concerns through the lens of US anti-discrimination law's disparate-impact doctrine and conclude that ``addressing the  sources of this unintentional discrimination and remedying the  corresponding  deficiencies in  the  law  will be difficult technically, difficult  legally, and difficult  politically.''

\underline{Value of \&}: By examining, in a manner that is both technically and legally rigorous, candidate approaches to cleansing data sets of past intentional discrimination, the authors expose ``practical  limits to what can be accomplished computationally.'' They state that a root-and-branch rethink of the meanings of ``discrimination'' and ``fairness'' may be needed in order to fix the problem of big data's disparate impact, and indeed a great deal of fundamental research into how to define and apply these concepts has been done since the paper was published nearly a decade ago.
\\

\item[$\circ$] 
\underline{Paper}: ``Using Zero-Knowledge to Reconcile Law Enforcement Secrecy and Fair Trial Rights in Criminal Cases,'' by Dor Bitan, Ran Canetti, Shafi Goldwasser, Rebecca Wexler~\cite{Bitan2022}

\underline{Relevant technology or application}: Peer-to-peer systems that require new investigative techniques, {\it e.g.}, apps with hidden back doors

\underline{Relevant law(s) or regulation(s)}: US constitutional right to ``confront evidence,'' {\it i.e.}, the right of the accused to see what his accuser has against him in order to assess its reliability.  This right derives from the Fifth and Sixth Amendments to the constitution.  

\underline{Summary}: The authors show how to augment peer-to-peer investigative software with a zero-knowledge proof mechanism in order to allow law-enforcement agencies to give full responses to challenges made by a defense expert while keeping the software hidden.  In this way, the authors are able to reconcile the need for law-enforcement secrecy with the right of a criminal defendant to inspect and challenge the evidence against him, including law enforcement’s investigative methods.

\underline{Value of \&}: By illustrating how zero-knowledge proof systems can be used to verify certain properties of software used in an investigation, this paper shows that scholarship at the intersection of computer science and law expands our understanding of how to apply established legal requirements to new technical contexts with the help of new system design.
\\

\item[$\circ$]
\underline{Paper}: ``Can the Government Compel Decryption? Don't Trust\ --\ Verify,'' by Aloni Cohen, Sarah Scheffler, and Mayank Varia~\cite{Cohen2022}

\underline{Relevant technology}: Interactive proof systems and cryptographic theory

\underline{Relevant law(s) or regulations(s)}: US constitutional right against self-incrimination, which is guaranteed by the Fifth Amendment to the constitution

\underline{Summary}: In light of the Fifth Amendment to the US constitution, a prosecutor cannot compel a defendant to perform an action that, by its very performance, would reveal implicit testimony. But this protection is not absolute: An action that communicates implicit testimony can be compelled if all of the implicit testimony is a foregone conclusion.  If the prosecutor already knows it, the implicit testimony would reveal little or nothing about the contents of the defendant’s mind, and the prosecutor would not be “relying on the truthtelling” of the defendant. Legal scholars refer to this principle as the {\it foregone-conclusion doctrine}.  Building upon prior work that applies zero-knowledge proof systems to the problem of compelled disclosure, the authors present an extensive and usable treatment of the foregone-conclusion doctrine that encompasses compelled actions, including authentication, commitment, hashing, and other “acts of production.”   

\underline{Value of \&}: This paper includes an important caveat about what can be discovered in the ``and'' between computer science and law: "We don’t want to mechanize the law. Law is flexible and adapts to new situations, and attempts to encode society’s rules into a rigid computer algorithm (aka, “code as law”) typically don’t end well. Hence, our pursuit of a computer science specification is not meant to replace the law, but to illuminate it."
\\

\end{enumerate}
\end{enumerate}

Of course, not all papers that satisfy the intersection rule fall into either of the two categories above.  We now present two examples of papers that don't.
\begin{enumerate}
\item[$\circ$]
\underline{Paper}: ``Non-Determinism and the Lawlessness of Machine Learning Code,'' by A.~Feder Cooper, Jonathan Frankle, and Christopher De Sa~\cite{Cooper2022}

\underline{Relevant technology}: Stochastic and non-deterministic computing

\underline{Summary}: The authors study error rates of ML systems and show that it is crucial to conceptualize the goals of such systems using the proper mathematical framework.  They analyze the essential stochasticity and non-determinism of ML systems and conclude that the goals should be conceptualized as mapping queries to sets of possible outputs or mapping training data to sets of possible models – not as mapping queries to outputs or training data to models in a deterministic, functional manner, as much earlier work had done.  

\underline{Value of \&}: By illuminating the important role of non-determinism, this work demonstrates that ML code falls outside of the cyberlaw frame of treating “code as law,” which assumes that code is deterministic. More generally, it demonstrates that scholarly efforts to align legal decision making with deterministic mathematical functions should be broadened to include stochasticity and non-determinism.
\\

\item[$\circ$] 
\underline{Paper}: ``Murmurs of the Silenced: Secure Reporting of Misconduct Settlements,'' by Peter K.~Chan, Alyson Carrel, Mayank Varia, and Xiao Wang~\cite{Chan2025}

\underline{Relevant technology}: Privacy-preserving computation

\underline{Summary}: The authors analyze misconduct settlements in five areas (employment discrimination, environmental pollution, housing discrimination, police misconduct, and sexual misconduct) in order to develop a statutory-technological system that implements secure reporting of wrongful misconduct settlements, thus balancing privacy with oversight.  The system uses secure, multi-party computation (MPC) in an essential way.

\underline{Value of \&}: By combining provably secure cryptographic-computing technology with rigorous legal analysis, the authors construct a statutory-technological system in which parties can provide private data for computation without giving up confidentiality.  The system enables secure reporting of wrongful misconduct settlements in order to provide oversight statistics to policymakers and to unmask repeatedly-settling parties for investigation.  It thus realizes a new point in the solution space of the longstanding debate on the merits between resolving disputes by public adjudications and resolving them by private settlements.
\end{enumerate}
We close this section by noting that there are no methodological restrictions in the CS\&Law scope.  Any mode of analysis or research methodology, be it theoretical or experimental, that is used by computer scientists or legal scholars is appropriate for a CS\&Law paper if it leads to interesting results that satisfy the intersection rule. Among the methodologies used in the examples given above are mathematical modeling, cryptographic theory, legal anthropology, and case studies. As the CS\&Law community grows and its intellectual scope evolves, it may invent new methodological approaches.  These inventions would be deployed along with the existing methodologies that the community has inherited from computer science and from law.

\section{Goals and Opportunities for Impact in Computer Science and Law Research}\label{section-non-research}

Computer Science and Law scholarship has high potential for impact on public policy and on legal and technical decisions made by responsible actors around the world. We describe here two types of contributions that are already evident in the field. First, CS\&Law research provides general guidance to policymakers, judges, and system designers as they make decisions about the framing of new law, the interpretation or application of existing law, and the design of systems meant to comply with various laws or internally defined rules. Second, in some cases, research in this field can make claims that are technically precise and aimed at testable propositions about whether laws meet their stated requirements or systems can claim to comply with laws. 

Many papers already discussed in Section~\ref{section-illustrative} provide examples of this first category of generalized guidance. Policy makers will benefit from Barocas and Selbst's observations about the need to reevaluate legal approaches to bias-based challenges they identify in current definitions of disparate impact~\cite{Barocas2016}. Scheffler's analysis of the privacy risks of age verification have informed an active debate in legislatures and before the US Supreme Court \cite{Scheffler2024}. (The fact that the US Supreme Court recently upheld a Texas age-verification law despite these technically-well-informed warnings suggests that CS\&Law researchers will have to be willing to stay closely involved in ongoing debates and recognize that guidance from the field may not always be accepted uncritically.\footnote{Free Speech Coalition, Inc. v. Paxton, 145 S. Ct. 2291 (2025)}) Judges will find Cohen, Scheffler, and Varia's work on compelled decryption useful in analyzing constitutional arguments as to when defendants will be required to decrypt data in a variety of system contexts~\cite{Cohen2022}. And courts adjudicating copyright claims may rely on Lee, Cooper, and Grimmelmann's description of generative-AI workflow in the use of arguably copyrightable subject matter~\cite{Lee2024}. 

Some CS\&Law research will offer claims stronger than the generalized  guidance we discuss above. Close examination of the relationship between a technical system and a law or regulation may analyze whether the law in question satisfies the goals it sets out. Or, from the technological perspective, it may ask how one can demonstrate that the system complies with the law.

For an example of an analysis of whether a law meets its goals, consider Meding and Sorge's work on Deep Fakes~\cite{Meding2025}, which shows that the EU AI Act employs inexact definitions of behaviors the law seeks to control, raising the risk that the law will not control harmful behaviors of certain AI models intended to be regulated. An early piece of CS\&Law scholarship on the GDPR and privacy explores how to demonstrate that a system comforms with a particular definition of identifiability in the law. Cohen and Nissim's proofs about Predicate Singling Out security~\cite{Cohen2020} provide a strong argument that any system meeting the formal requirement they describe, all other things being equal, provably satisfies one particular aspect of the GDPR. Such strong claims as these make a valuable contribution to greater public confidence in the application of law to computer systems and decrease the burden of compliance by increasing certainty about the precise requirements of legal rules.

\section{Future Directions}\label{section-future}

\subsection{Intellectual Challenges\label{subsec-intellectual}}

The intellectual future of CS\&Law is exceptionally promising.  None of the sociotechnical issues with which our community engages is fully resolved; over the forseeable future, all will continue to be important, and some will become more important.  These issues include AI regulation, online privacy, protection of digital intellectual property, freedom of expression online, and other open-ended questions that have been partially, but not fully, answered by papers in CS\&Law and adjacent venues.  We briefly describe two topics on which there is opportunity for highly impactful work; they are two of many.

Generative AI systems have been (and are still being) trained on vast amounts of material, including but not limited to material that is protected by copyright, trademark, and other types of intellectual property law.  The recent settlement of a class-action lawsuit, brought by three authors who claimed that Anthropic built its business on ``large-scale copyright theft,'' is evidence that the generative-AI industry and the courts accept the proposition that creators whose intellectual property is used for training deserve compensation.  However, Anthropic's ability to pay \$1.5 billion (\$3K per book) does not provide a widely applicable model.  How should individuals and small companies that train models using modest amounts of others' intellectual property compensate the IP owners?  How can IP owners be identified, reasonable prices be determined without lengthy and expensive negotiations, and payments delivered?

Freedom of expression online is a cherished value in the CS\&Law community (or at least most of it).  However, wide circulation of harmful misinformation is a serious problem that threatens the physical safety, psychological well being, and economic flourishing of people everywhere.  Efforts to label and/or prevent the dissemination of misinformation during the COVID-19 pandemic were not particularly successful and created a backlash the political ramifications of which are still being felt.  Some are tempted to give up, but the threat is serious enough to warrant the attention of the CS\&Law community.  How can technology and law work together not to deny people the freedom to post misinformation online but rather to blunt the harmful effects of such posts?

\subsection{Funding Challenges}\label{subsec-funding}

Unfortunately, the retreat from the endless frontier\footnote{In July 1945, Vannevar Bush submitted to President Harry S. Truman his report entitled “Science: The Endless Frontier.” The National Science Foundation, which Bush had envisioned, was founded in 1950.  During that five-year gap, other federal agencies, including the Atomic Energy Commission, the National Institutes of Health, and the Office of Naval Research, expanded their scientific-research programs. Thus began the roughly 80-year golden age of US-government support for basic and applied research that is now winding down.} is real, and it has touched computer science and all of its subdisciplines. The CS\&Law community cannot, therefore, count on being able to obtain US federal grants with a reasonable amount of time and effort, as it could when the symposium series was founded in 2019. US federal funding will not entirely disappear overnight, however, and there may be a silver lining for relatively young research areas if the established funding agencies need to justify their existence anew or states step up to create scientific-research programs.  

For example, many people, including some in the current US administration that has evinced such deep skepticism about university-based research, believe that generative AI changes everything and/or that the US must contest China for global dominance in AI.  They therefore have good reason to support research that studies generative AI from a CS\&Law perspective.  Even anti-regulation AI boosters want laws that promote their technology and their companies.  The CS\&Law community should take every opportunity it has to explain why both the laws and the technology should be crafted by people who follow the intersection rule and that the best way to nurture a community of such people is to fund its research.

The tech industry, in particular the AI industry, may also be a crucial funding source going forward. Past CS\&Law conferences have included authors and panelists who work in industry as well as university-based authors whose research was funded by industrial grants or gifts.  They have been few in number so far, but perhaps that could change if companies see the US government's retreat from the endless frontier as a threat to their talent pool and the basic research that they build upon.

Of course CS\&Law is an international conference, and government-funding agencies throughout the world have supported research that was reported in CS\&Law papers.  We hope that those funding streams will continue and expand.

\section*{Acknowledgements}
Joan Feigenbaum holds concurrent appointments as the Grace Murray Hopper Professor of Computer Science at Yale University and as an Amazon Scholar. This paper describes work performed at Yale and is not associated with Amazon. 

Daniel Weitzner is the 3Com Founders Senior Research Scientist at the MIT Computer Science and Artificial Intelligence Lab and the Founding Director of the MIT Internet Policy Research Initiative. His work was funded in part by NSF grant CCF-2131541 (DASS: Legally Accountable Cryptographic Computing Systems (LAChS)) and the MIT Future of Data Initiative.

\bibliographystyle{plain}
\bibliography{SoK}

\begin{thebibliography}{10}

\bibitem{Abelson1997}
Hal Abelson, Ross Anderson, Steven~M. Bellovin, Josh Benaloh, Matt Blaze, Whitfield Diffie, John Gilmore, Peter~G. Neumann, Ronald~L. Rivest, Jeffrey~I. Schiller, and Bruce Schneier.
\newblock The risks of key recovery, key escrow, and trusted third-party encryption, May 1997.

\bibitem{Abelson2015}
Harold Abelson, Ross Anderson, Steven~M. Bellovin, Josh Benaloh, Matt Blaze, Whitfield Diffie, John Gilmore, Matthew Green, Susan Landau, Peter~G. Neumann, Ronald~L. Rivest, Jeffrey~I. Schiller, Bruce Schneier, Michael~A. Specter, and Daniel~J. Weitzner.
\newblock Keys under doormats: mandating insecurity by requiring government access to all data and communications.
\newblock {\em Journal of Cybersecurity}, 1(1):69--79, September 2015.

\bibitem{Barocas2016}
Solon Barocas and Andrew~D. Selbst.
\newblock Big data's disparate impact.
\newblock {\em California Law Review}, 104:671--732, 2016.

\bibitem{Bitan2022}
Dor Bitan, Ran Canetti, Shafi Goldwasser, and Rebecca Wexler.
\newblock Using zero-knowledge to reconcile law enforcement secrecy and fair trial rights in criminal cases.
\newblock In {\em Proceedings of the 2022 Symposium on Computer Science and Law}, page 9–22, New York, 2022. ACM.

\bibitem{Burmeister2021}
Fabian Burmeister, Christian Kurtz, Pascal Vogel, Paul Drews, and Ingrid Schirmer.
\newblock Unraveling privacy concerns in complex data ecosystems with architectural thinking.
\newblock In {\em Proceedings of the $42^{nd}$ International Conference on Information Systems}, pages 1--17, Atlanta, 2021. Association for Information Systems.

\bibitem{Burmeister2022}
Fabian Burmeister, Mickey Zar, Tilo B{\"o}hmann, Niva Elkin-Koren, Christian Kurtz, and Wolfgang Schulz.
\newblock Toward architecture-driven interdisciplinary research: Learnings from a case study of covid-19 contact tracing apps.
\newblock In {\em Proceedings of the 2022 Symposium on Computer Science and Law}, page 143–154, New York, 2022. ACM.

\bibitem{Chan2025}
Peter~K. Chan, Alyson Carrel, Mayank Varia, and Xiao Wang.
\newblock Murmurs of the silenced: Secure reporting of misconduct settlements.
\newblock In {\em Proceedings of the 2025 Symposium on Computer Science and Law}, pages 121--135, New York, 2025. ACM.

\bibitem{Cohen2020}
Aloni Cohen and Kobbi Nissim.
\newblock Towards formalizing the gdpr’s notion of singling out.
\newblock {\em Proceedings of the National Academy of Sciences}, 117(15):8344–8352, April 2020.

\bibitem{Cohen2022}
Aloni Cohen, Sarah Scheffler, and Mayank Varia.
\newblock Can the government compel decryption? don't trust\ --\ verify.
\newblock In {\em Proceedings of the 2022 Symposium on Computer Science and Law}, pages 23--36, New York, 2022. ACM.

\bibitem{Cooper2022}
A.~Feder Cooper, Jonathan Frankle, and Christopher~De Sa.
\newblock Non-determinism and the lawlessness of machine learning code.
\newblock In {\em Proceedings of the 2022 Symposium on Computer Science and Law}, pages 1--8, New York, 2022. ACM.

\bibitem{Feigenbaum2018}
Joan Feigenbaum and Daniel~J. Weitzner.
\newblock On the incommensurability of laws and technical mechanisms: Or, what cryptography can’t do.
\newblock In {\em Proceedings of Security Protocols XXVI}, volume 11286 of {\em Lecture Notes in Computer Science}, page 266–279, Cham, Switzerland, 2018. Springer.

\bibitem{Lee2024}
Katherine Lee, A.~Feder Cooper, and James Grimmelmann.
\newblock Talkin’ ‘bout {AI} generation: Copyright and the generative-ai supply chain (the short version).
\newblock In {\em Proceedings of the 2024 Symposium on Computer Science and Law}, pages 48--63, New York, 2024. ACM.

\bibitem{Meding2025}
Kristof Meding and Christoph Sorge.
\newblock What constitutes a deep fake? the blurry line between legitimate processing and manipulation under the eu ai act.
\newblock In {\em Proceedings of the 2025 Symposium on Computer Science and Law}, pages 37--49, New York, 2025. ACM.

\bibitem{Samuelson1994}
Pamela Samuelson, Randall Davis, Mitchell~D. Kapoor, and J.~H. Reichman.
\newblock A manifesto concerning the legal protection of computer programs.
\newblock {\em Columbia Law Review}, 94(8):2308--2431, December 1994.

\bibitem{Scheffler2024}
Sarah Scheffler.
\newblock Age verification systems will be a personal identifiable information nightmare.
\newblock {\em Communications of the ACM}, 85(7):31–33, July 2024.

\end{thebibliography}
\end{document}